# Crossover from ‘mesoscopic’ to ‘universal’ Phase for electron transmission in Quantum Dots

M. Avinun-Kalish, M. Heiblum, O. Zarchin, D. Mahalu & V. Umansky

*Braun Center for Submicron Research, Department of Condensed Matter Physics,*

*Weizmann Institute of Science, Rehovot 76100, Israel*

**Measuring phase in coherent electron systems (‘mesoscopic’ systems) provides ample information that not easily revealed by conductance measurements. Phase measurements in relatively large quantum dots (QDs) [1] recently demonstrated a ‘universal’ phase evolution independent of dot size, shape, and occupancy [2, 3]. Explicitly, in Coulomb blockaded QDs the transmission phase increased monotonically by π throughout each conductance peak, thereafter, in the conductance valleys the phase returned sharply to its base value. Expected ‘mesoscopic’ features in the phase, related to dot’s shape, spin degeneracy or to exchange effects, were never observed. Presently, there is no satisfactory explanation for the observed phase ‘universality’ [4]. Unfortunately, the phase in a few-electron QDs, where it can be better understood, was never measured. Here we report the results of such measurements performed on a small QDs that occupy only 1-20 electrons. Unlike the repetitive behavior found in larger dots we found ‘mesoscopic’ features for dot occupation of less than some 10 electrons. As the occupation increased the phase evolved and turned ‘universal’ for some 14 electrons and more. Aside from the detailed phase behavior of a few electron dots, these measurements might help in singling out the correct theoretical model for the ‘universal’ behavior.**

The actual experimental configuration is described in Fig. 1. The interferometer, formed in a two dimensional electron gas (a GaAs-AlGaAs heterostructure), consists of emitter $E$ and collector $C$ constrictions, each formed by a quantum point contact (QPC), and a few base regions $B$ in between. The grounded base regions (chemical potential $\mu_B = 0$) serve as draining reservoirs, ensuring that interference is only between two paths [5]. A center island separates the incoming electrons into two paths and embeds within it the plunger gate of the QD, which controls the occupancy of the dot. The transmission phase of the QD adds to the phase gain in the left arm and is then compared with the phase of the right arm. Threading a weak magnetic field through the area of the interferometer adds an Aharonov-Bohm (AB) phase to one arm $\Delta\varphi_{AB} = 2\pi\,\Phi/\Phi_0$, where $\Phi$ is the magnetic flux enclosed by the two arms and $\Phi_0$=$h/e$ is the flux quantum [6,7]. A QPC is coupled capacitively to the QD, hence sensing its potential – enabling counting the electrons in the dot [8,9].

A simplified model of the QD is a resonant tunnelling device: a potential well confined between two tunnelling barriers with quasi bound resonant states $E_n$. The transmission amplitude exhibits maxima whenever the Fermi energy in the leads $E_F$ coincides with $E_n$. Due to the finite coupling between well and leads, the levels are broadened to $\Gamma_n$ (dwell time ~$h/\Gamma_n$), with the transmission described by the Breit-Wigner expression [10] $t_{QD} = C_n \frac{i\Gamma_n/2}{E_F - E_n + i\Gamma_n/2}$.

Here, the $C_n$'s are positive or negative real numbers (since at a zero magnetic field the Hamiltonian is real), depending on the parity of the $n^{\text{th}}$ wave function with respect to the 'in' and 'out' openings of the QD. The phase of the transmission amplitude evolves by $\pi$ through each resonance while the relative phase between adjacent resonances, determined by $C_n$, can be 0 or $\pi$ [11]. However, in contrast, all previous measurements, conducted with relatively large occupancy QDs, consistently led only to positive $C_n$'s [2, 3, our unpublished data].

A large number of theoretical publications had been devoted to the puzzle of phase evolution (see recent summary by Hackenbroich [4]). They may be grouped in three main classes. The first class questions whether the measured phase is the 'intrinsic transmission phase' of the QD or a modified phase due to multiple paths traversing the interferometer [12, 13]. The second class considers transport that is mediated by interplay of more than one quantum state. A common scenario assumes an existence of a dominant level strongly coupled to the leads, responsible to shuttle the electrons [14, 15]. After occupation the electron is unloaded to a localized level, weakly coupled to the leads, allowing the dominant level to be free again to transfer another electron. Hence, the observed phase is only that of the dominant level. Based on this idea other models examined only two levels, with one of the levels dominant, adding spins, adding interactions, or assuming a finite temperature [16, 17, 18]. Interaction between two levels was invoked also in a QD where the plunger gate couples with different strengths to different energy states, leading thus to avoided level crossing and charge shuttling between levels [4]. The third class deals with specific energies where both the imaginary and the real parts of the transmission coefficient vanish [19]. These singular points, that explain the phase slips in the valleys, might result due to a deviation from a strictly 0D system [4] or to an existence of Fano resonances in the dot [20]; but can not explain the 'in phase' behavior of all peaks. Naturally, one would expect the breakdown of every model for some tuning parameters, which we, thus far, never observed. Still, some models may predict an 'in phase' behavior for a very large sequence of peaks. We return to this issue later.

At low enough temperatures both the phase coherence length and the elastic mean free path exceed the sample size. The current in a grounded collector is $I_C = \frac{2e^2}{h} T_{EC} V_{EB}$, where $V_{EB}$ is the voltage applied between emitter and base and $T_{EC}$ is the transmission probability from emitter to collector [21]. The transmission $T_{EC}$ results from a coherent sum of the two trajectory amplitudes that traverse the two arms of the interferometer, $T_{EC} = |t_{EC}|^2$ and $t_{EC}=t_L+t_R$, where $t_{L,R}$ the transmission amplitudes associated with the left and right paths, respectively. The phase dependent part is $\tilde{T}_{EC} \propto |t_L||t_R| \cos[\Delta\varphi_{AB} + \varphi(t_R) - \varphi(t_L)]$, with $\varphi\ (t_{L,R})$ the corresponding phases, and $\varphi(t_L)=\varphi_0(t_L)+\varphi_{QD}$ the accumulated phase in the left arm. Hence, $\tilde{T}_{EC}$ oscillates as a function of magnetic flux with period $\Phi_0$, with any change in $\varphi_{QD}$ leading to a similar change in the phase of the oscillating collector current.

Measurements were done in a dilution refrigerator ($T_{lattice} \approx 20\,mK$, $T_{electrons} \approx 30\,mK$) with an AC excitation voltage $V_{EB} = 2-20\,\mu V$ (frequency ~23 Hz). The integrity of the 'two-path' interferometer was verified by observing a single period $\Delta\Phi=\Phi_0=h/e$ in the interference signal (see Fig. 3), indicating only two path interference. Higher orders, with period $\Phi_0/n$, were at least 4 orders of magnitude smaller. The QD was formed by adjusting the resistance of its 'in' and 'out' QPCs to be greater than $h/2e^2$, namely, in the Coulomb blockade (CB) regime. To ensure transport mainly through one level we tuned the dot to $\Gamma < \Delta E$, with $\Delta E$~0.5 meV and $\Gamma$=30-300 μeV, with the temperature being the smallest energy, $k_BT$<3 μeV. In the 'constant interaction' model the complex interaction among the electrons is represented by a capacitor $C_{QD}$, leading to a charging energy $U=e/2C_{QD}$ ~ 1-3 meV. Varying the plunger gate voltage $V_P$ changes the potential landscape in the QD and consequently the occupation $N$. Classically, for a certain $V_P$ degeneracy takes place, $E(N)=E(N+1)$; allowing the number of electrons to fluctuate between $N$ and $N+1$ with no energy cost, allowing current flow.

The electron counting detector is a separately biased QPC, capacitively coupled to the QD [8,9]. The induced potential in the QPC is $V_{QPC} = V_{QD}\frac{C_{QPC-QD}}{C_{QPC}}$, where $C_{QPC}$ is the self capacitance of the QPC and $C_{QPC\text{-}QD}$ is the mutual capacitance between QD and QPC detector. Charging the QD affects $V_{QPC}$ and the conductance of the QPC, which in turn is analyzed by a small current (20-80 nA). The potential energy of the QD rises linearly with plunger voltage, reaching eventually $\Delta E+e^2/2C_{QD}$, followed by a sharp drop when an electron enters the dot. Indeed, the conductance of the QPC detector exhibited a repetitive 'saw-tooth' like oscillations as a function of $V_P$, with one period for every additional electron entering the dot. We measured the derivative $dI_{QPC}/dV_P$ (via applying a small AC voltage to the plunger gate), resulting with easily identifiable dips in the derivative (see Fig. 2). Note that determining the occupancy of the QD via a separate detector is necessary since for sufficiently negative $V_P$ the QD inadvertently pinches off (via the mutual coupling among the different gates). Consequently, the dot's conductance peaks weaken and are impossible to resolve. As seen in Fig. 2a, the conductance dips of the detector persisted down to $V_P\approx$-400 mV, much beyond $V_P\approx$-300 mV where the conductance peaks of the QD were too small to resolve. We carefully retuned the QD without changing its occupancy, in order to maximize the conductance peaks, allowing reliable phase measurements (e.g. Fig. 2b).

Varying the magnetic field in a range 0-30 mT, after the QD was tuned to conduct, resulted in relatively high visibility AB oscillations with a single period $\Delta B=\Phi_0/\text{area} \cong 2$ mT (see Fig. 3; in a different structure $\Delta B$ ~3.5 mT). The coherent part of the transmission (Fig. 3b) and the transmission phase (Figs. 4-6) were both determined from data such as shown in Fig. 3a by performing a fast complex Fourier transform of the AB oscillations as a function of $V_P$.

We studied two different configurations of interferometers and QDs, as well as on thermally recycled structures (which behave as different devices after thermal recycling). A single tuning allowed the addition of only 2-3 electrons without changing drastically the coupling of the dot to the leads or the symmetry of the interferometer. Hence, the QD and the interferometer were retuned after every few added electrons, keeping the occupation at check with the QPC detector, in order to optimize the visibility and CB conditions. The measured phase in different occupation regimes was then patched together in order to obtain a continuous phase evolution over a wide range of electron occupation. We present in Figs. 4-6 examples of phase and amplitude of the coherent part of the transmission coefficient for an increasing electron number in the QD. We did not subtract any extraneous phase that may be induced in the arms of the interferometer by the varied plunger voltage since this phase is difficult to determine accurately. However, we estimated it to be a weak function of $V_P$ and quite small for the addition of 2-3 electrons at a time.

Already the phase evolution across the first two conductance peaks and valleys (the first two electrons entering the dot) exhibited a marked deviation from the 'universal' behavior. As demonstrated in the results of two different dot & interferometer designs, the phase climbed by $\pi$ for each of the first two added electrons (Fig. 4). Moreover, this dependence was robust and independent of dot's tunings. Evidently, the different phase of the first two conductance peaks suggest that the first two electrons occupy two opposite parity orbital states, and not one state as assumed thus far [22, 23, 24]). Since the ground state of a 'two-electron-system' must be a 'singlet' [25], two opposite spins occupy the lowest two states. This is not surprising since our dots are likely to have a very shallow potential well, hence a relatively small single particle level spacing, favoring an occupation of two levels in order to minimize the Coulomb repulsion [26]. We will not speculate here on the reproducible dip in the phase prior to the entering of the second electron, which is as large as $\pi/2$ (see Fig. 4b). Adding the third and fourth electrons (Fig. 4a), the phase evolves from $\pi$ to $2\pi$ - a similar range of the second electron. Consequently, this data suggest that the second through the fourth electrons all occupy similar parity orbital levels. Similarly, the fifth electron evolves from $\sim 2\pi$ to $\sim 3\pi$, which is indistinguishable from a phase evolving from 0 to $\pi$ of the first electron.

We devote Figs. 5a and 5b to show the sensitivity of the phase to the tuning parameters. While the phase in the fifth, sixth and seventh conductance peaks is independent of tuning parameters, the phase of the eighth electron depends on the tuning parameters. For somewhat different parameters of the QD the eighth electron has a different than that of the seventh electron. The above described examples (Figs. 4 and 5) clearly demonstrate a reasonable phase behavior, sensitive to details of the potential, showing spin degeneracy, exchange interaction, or Cooper channel interaction [26]. This is distinctly different from the repetitive, 'universal' like, phase rigidity observed in many electron dots [2, 3].

With the addition of more electrons the phase evolves through a transition region, which resembles the 'universal' behavior (not shown). The phase climbs throughout each conductance peak and drops in the valleys, riding though on a rising background phase. However, for dot occupation of some 14 electrons and higher (Fig. 6) the phase 'locks' into the 'universal' behavior - being then insensitive to dot's tuning! In other words, the intricacies of shape dependent level occupation and parity effects disappear altogether.

Since our aim was to explore the validity of the bizarre phase evolution over a large range of parameters, we may now ask: what have we learnt from this new set of measurements? First, that there are two distinct regimes of phase evolution. (a) For occupation 1 to about 10; a highly sensitive phase to dot's configuration and occupation. (b) For occupation higher than about 14; phase evolution is 'universal' like, with phase ranging only between 0 and $\pi$, and is independent of dot's configuration and occupation. While regime (a) can be explained by current understanding of QDs, regime (b) presents difficulties, especially since screening is expected to be more effective at higher occupations (possibly leading to single particle like behavior). However, the absence of phase 'universality' in regime (a) is gratifying for two main reasons: First, it validates that our measured phase is indeed the intrinsic phase of the QD. Otherwise, one would not observe such distinct differences between dilute and highly populated dots. Second, we can now comment on some aspects of the existing theories. An outstanding feature of larger QDs is the smaller level spacing, which might lead to levels overlapping. This will favor models that invoke transmission mediated through more than one quantum state. An illuminating example can be that of a dominant strongly coupled orbital state to the leads. Such special level can be simply a solution of the Schrödinger equation [15], or alternatively, the higher between any two sequential levels. The higher is broader since it experiences a lower potential barrier to the leads [Y. Oreg, private communication]. Consequently, a broader level is likely to be occupied in a large range of energies, hence responsible for electron shuttling and the similar phase of many peaks. Still, a theory has to be developed that explains the robustness of the effect for an extremely large number of conductance peaks.

**Acknowledgement**

The work was partly supported by the Minerva foundation, the German Israeli Project Cooperation (DIP), the German Israeli Foundation (GIF), and the QUACS network. We are grateful to Y. Levinson, Y. Oreg, and A. Yacoby for useful discussions. We thank Milos Popadic for assistance in the lab.

Correspondence and requests for materials should be addressed to Moty Heiblum. (heiblum@wisemail.weizmann.ac.il)

## Figures

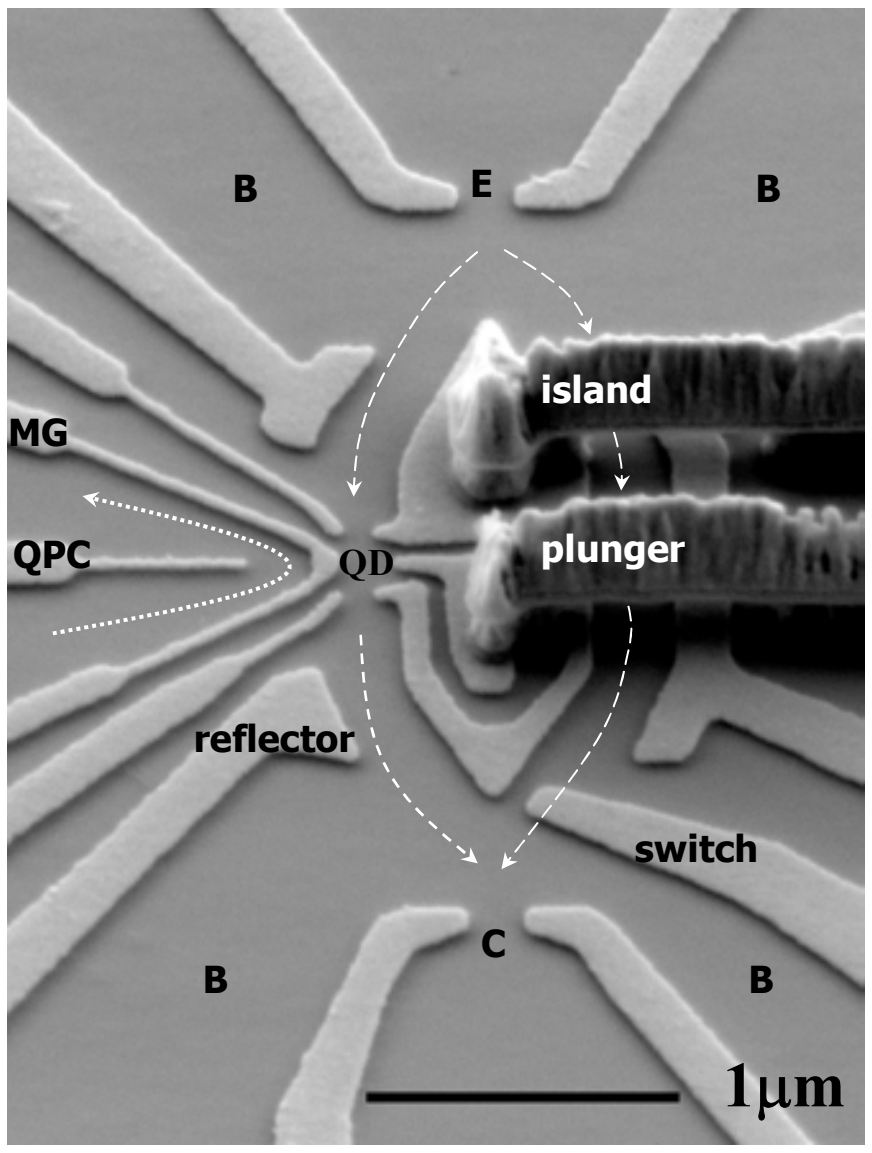

**Fig. 1.** SEM micrograph of the actual device. The actual *two path* interferometer consists of a patterned high mobility two dimensional electron gas formed some 53 nm below the surface of a GaAs-AlGaAs heterostructure, with areal carrier density $n_s = 3.3 \cdot 10^{11}\ cm^{-2}$ and mobility $\mu = 1.6 \cdot 10^6\ cm^2/Vs$ at $T = 4.2\ K$, resulting in an elastic mean free path $l = 14.5\ \mu m$. To assure only two interfering paths from *E* to *C*, the reflected and scattered paths are collected by the base *B* regions through openings formed between the gates of the reflectors and the QPCs The 'switch' gate turns off the right path to allow tuning of the QD. The QD is composed of two QPCs and a middle gate (MG) in between allowing forming 'small' and 'large' dots. The 'plunger' gate is embedded in the center island and is connected via a metallic air bridge to the outside in order to allow biasing without crossing the right (reference) arm of the interferometer. Similarly, another bridge biases the center island. The magnetic flux is contained in the area between the two paths (shown by the dashed lines). The QPC counting detector, separated from the QD by MG, has its own current path (shown by the dotted line). The transmission from *E* to *C* oscillates with the AB flux with phase dependent on the transmission phase of the QD.

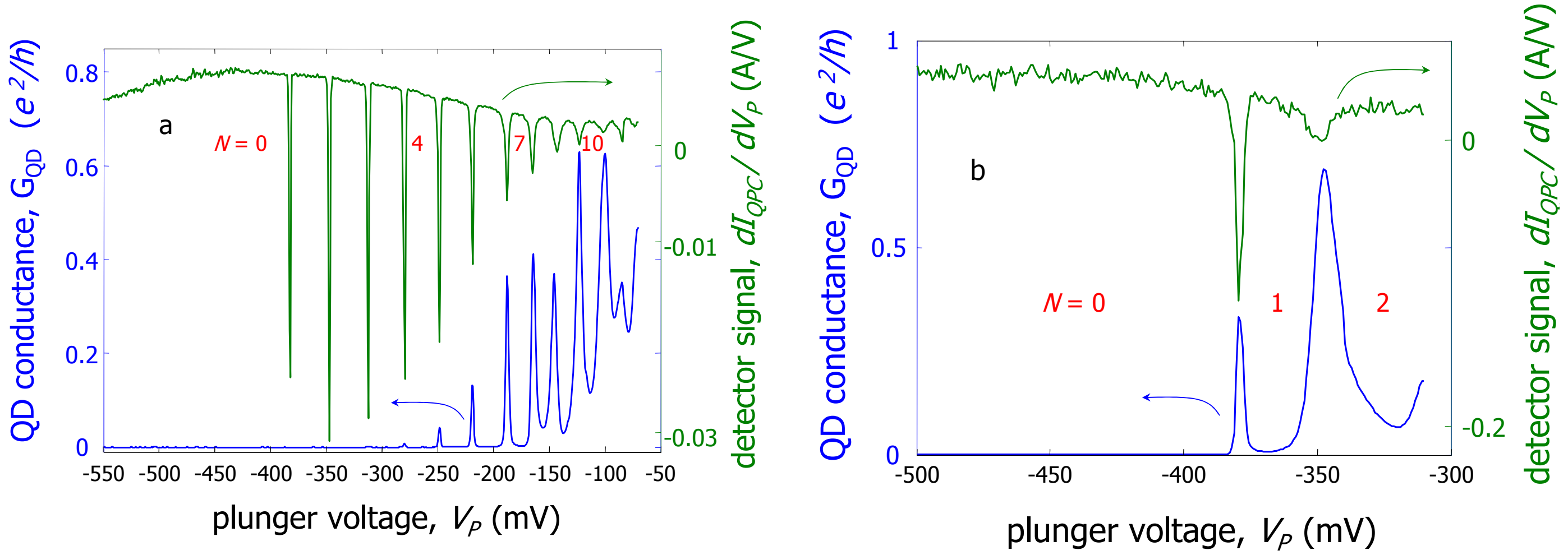

**Fig. 2.** Electron counting in the QD by a QPC detector. In the CB regime we expect $V_{QD}$ to increase as the plunger gate voltage $V_P$ increases, eventually reaching $V_{QD}=(e/2C_{QD})+\Delta E/e$, with $C_{QD}$ the self capacitance of the dot and $e/2C_{QD}$ the charging energy of the dot. At this potential an electron enters the QD and screens the positive potential induced by the plunger resulting in a potential drop. Hence, the QD potential, and consequently the conductance of the QPC counting detector, exhibit a saw-tooth like behavior. The derivative $dI_{QPC}/dV_P$ is a series of negative dips. (a) Conductance peaks of the QD (blue) and the corresponding detector dips (green). The detector proves the presence of electrons even though the QD conductance is too small to be measured. (b) The QD is retuned by opening the *in* and *out* QPCs and changing the voltage on MG in order to allow measurable conductance of the first few electrons.

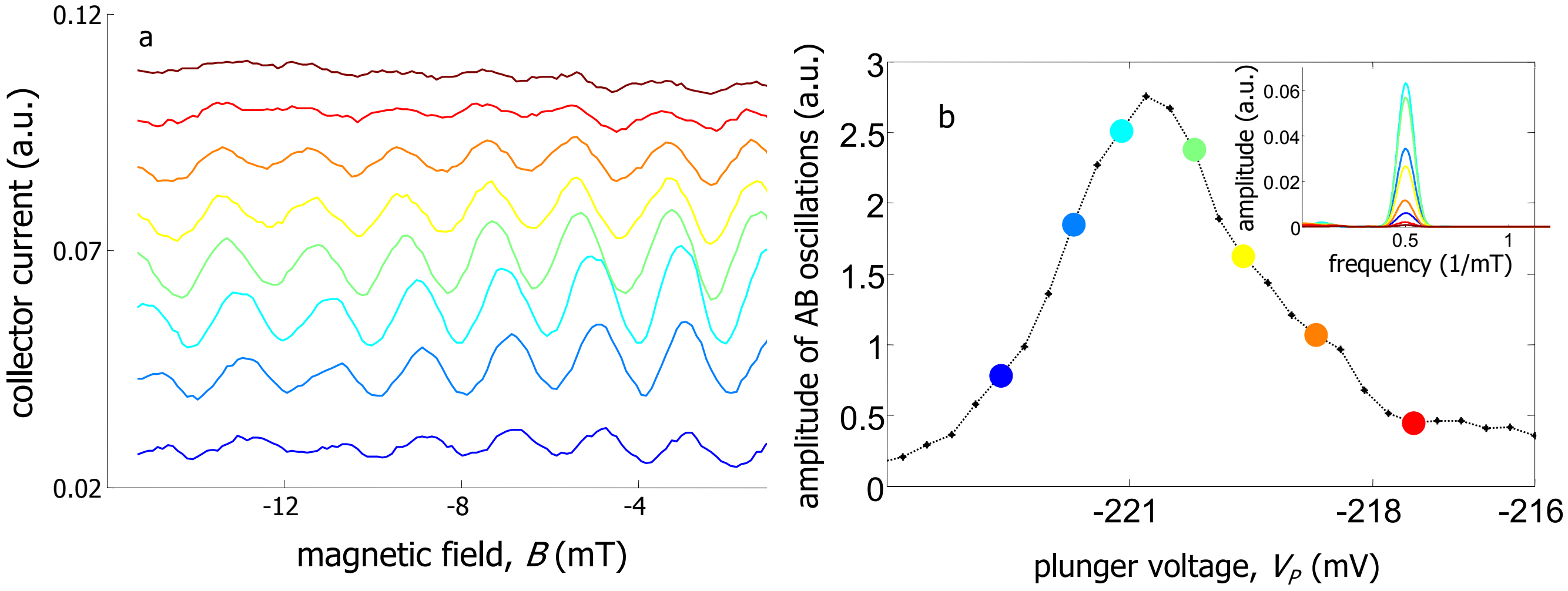

**Fig. 3.** Phase measurements procedure. (a) The QD is first tuned to conduct and the AB oscillations (with $V_P$ as a parameter) are monitored through out a conductance peak at the chosen values shown in Fig. 3b. Note the change in amplitude and in phase among the different traces (that are shifted vertically for clarity). (b) The amplitude of the oscillations plotted as a function of $V_P$ (the colored points correspond to the colored traces in (a)). The amplitude was determined by a fast Fourier transform. Insert in (b). Fourier transforms of the oscillations indicating a single AB period corresponding to an addition of a flux quantum *h/e* to the area enclosed by the two paths.

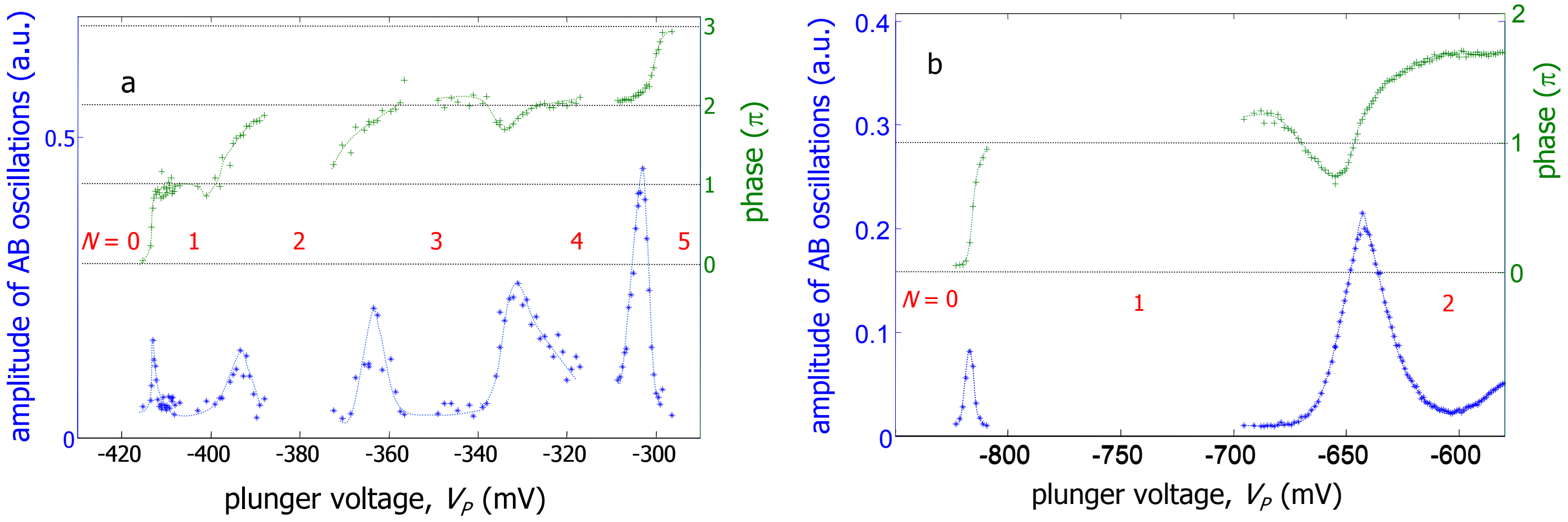

**Fig. 4.** Phase evolution and coherent conductance for the first few electrons in the QD. The dotted lines are just a guide to the eye. (a) Typical behavior of the phase for the first five electrons. While the first two electrons enter different orbital states the following two share the same orbital parity with the second electron. (b) Detailed behavior of the phase for the first two electrons in a different device. Note the two-orbital singlet, which is robust for all tuning parameters. Preceding entering of the second electron the phase exhibits always a drop – as large as π/2 – independent of how much the QD is being pinched. It may be related to an onset of Kondo correlation.

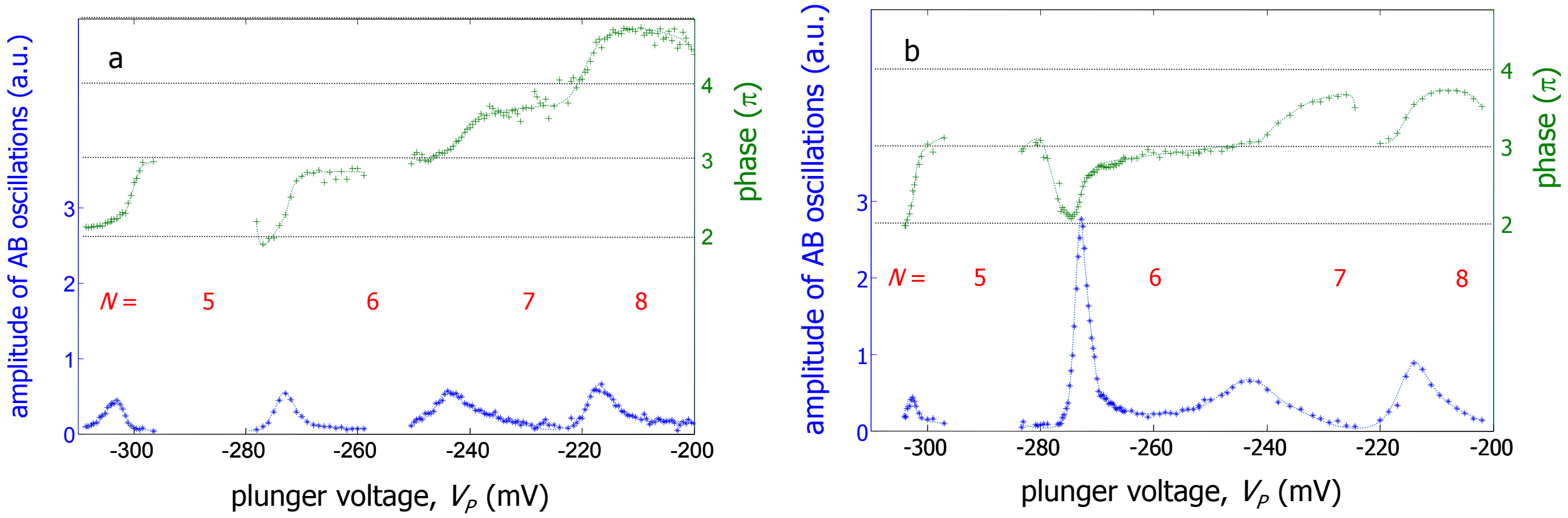

**Fig. 5.** Two examples of phase evolution for $N$=5-8 with different tuning parameters of the QD and the interferometer. This is to demonstrate that the features are mesoscopic, namely, sensitive to the dot's configuration. While the fifth and sixth electrons remain in similar parity states for the two tunings, the seventh and eighth electrons are in different parity orbital states in (a) and in similar parity orbital states in (b).

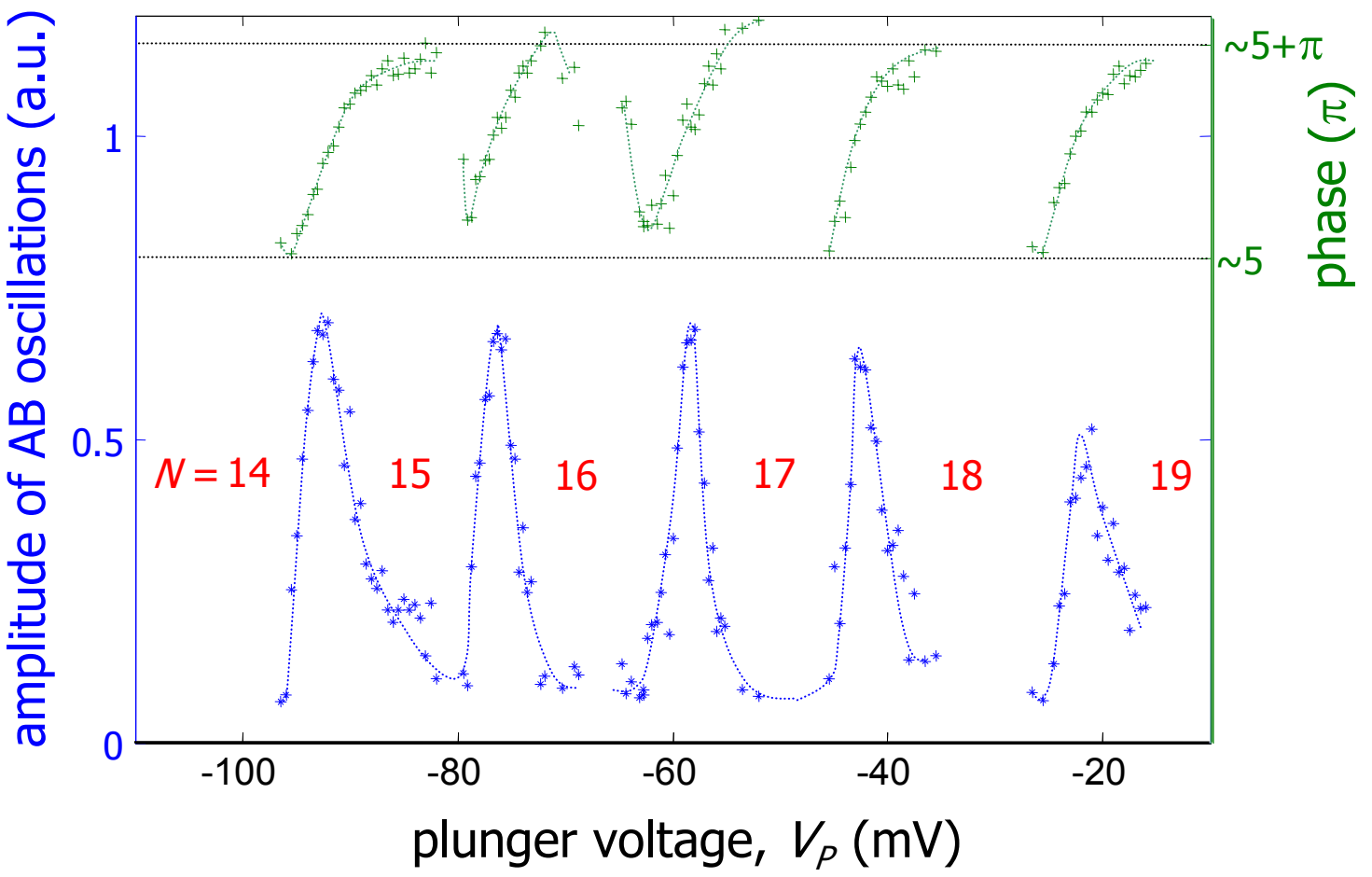

**Fig. 6.** A demonstration of the 'universal' phase evolution the dot enters after fourteen electrons. This behavior is independent of tuning and is ubiquitous to all measured *many-electron* dots. Note that the absolute value of phase (with respect to the phase of the first electron) is difficult to determine accurately due to an accumulated phase in the leads or a distortion of the interferometer, hence, only its approximate value is noted, however, the phase span is always π.